\documentclass[reprint,amsmath,amssymb,aps,pra]{revtex4-1}
\usepackage[utf8]{inputenc}
\usepackage{graphicx}% Include figure files
\usepackage{dcolumn}% Align table columns on decimal point
\usepackage{mathrsfs,amsmath}
\usepackage{bm}
\def\ph2{{\it p}-H$_2$}
\def\Am3{\AA$^{-3}$}

\begin{document}
\title{Specific heat of thin $^4$He films on graphite}
\author{Massimo Boninsegni}
\affiliation{Department of Physics, University of Alberta, Edmonton, Alberta, T6G 2E1, Can playada}
\email{m.boninsegni@ualberta.ca}
\author{Saverio Moroni}
\affiliation{CNR-IOM Democritos, Istituto Officina dei Materiali and Scuola Internazionale Superiore di Studi Avanzati,
Via Bonomea 265, I-34136 Trieste, Italy}
\date{\today}
\begin{abstract}
{
The specific heat of a two-layer $^4$He film adsorbed on a graphite substrate is estimated as a function of temperature by Quantum Monte Carlo simulations. The results are consistent with recent experimental observations [S. Nakamura {\em et al.}, Phys. Rev. B {\bf 94}, 180501(R) (2016)], in that they broadly reproduce their most important features. However, neither the ``supersolid'' nor the ``superfluid hexatic'' phases, of which experimental data are claimed to be evidence, are observed. It is contended that heat capacity measurements alone may not be a good predictor of structural and superfluid transitions in this system, as their interpretation is often  ambiguous.
}
\end{abstract}
\maketitle 

\section{Introduction}
\label{sec:intro}
% TODO: write your article here.
The experimental  investigation of the phase diagram of a thin (i.e., few layers) film of helium adsorbed on graphite has been pursued for half a century \cite{bretz,hering,polanco,carneiro,ecke,lauter,lauter2,greywall,greywall2}, motivated by the dazzling variety of phases that this system displays. In spite of such an impressive effort, the subject remains marred in controversy, mainly centered on the possible existence of a ``supersolid''\cite{notea} phase in the second adsorbed layer of $^4$He. \\ \indent 
It was first suggested by Greywall and Busch \cite{greywall,greywall2} that  a commensurate crystalline phase may exist in the second $^4$He adlayer, with a $\sqrt 7 \times \sqrt 7$ partial registry  with respect to the first layer (this phase is henceforth referred to as 4/7). Crowell and Reppy \cite{crowell,crowell2} subsequently proposed that such a phase may turn superfluid at sufficiently low temperature, making it a rare example of a phase of matter simultaneously displaying structural and superfluid order.
\\ \indent
The existence of the 4/7 commensurate phase is a {\em conjecture} put forth to account for heat capacity measurements whose interpretation is not univocal;  no {\em direct}, conclusive experimental evidence of such a phase has been reported to date. Theoretical studies based on first principle Quantum Monte Carlo (QMC) simulations, making use of realistic microscopic atom-atom and atom-surface potentials \cite{corboz,happacher,ahn,sm}, lend no support to the scenario of a 
commensurate solid in the phase diagram of the second layer; rather, the system is predicted to display only superfluid and incommensurate crystalline phases.  Assuming a two-dimensional (2D) first layer density between 0.118 and 0.122  \AA$^{-2}$ \cite{lauter2,fukuyama}, one ends up with a 2D upper layer density for the hypothetical 4/7 phase between 0.067 and 0.070 \AA$^{-2}$;  at, or near that density, a non-crystalline superfluid phase is predicted by the most reliable theoretical calculations, all the way to zero temperature, with a phase transition between a superfluid and an {\em incommensurate} crystal taking place at a $\sim 10$\% higher second layer density.\\ \indent
However, recent heat capacity measurements \cite{fukuyama} have again been interpreted as signalling the occurrence of a {\em  commensurate} (not 4/7) ``supersolid'' phase. Specifically, an observed broad peak in the specific heat of a $^4$He film, reaching its maximum height at a temperature $\sim 1.4$ K for a two-layer film of total coverage  $\theta_A=0.1973$ \AA$^{-2}$ (a very similar peak was  observed in previous work \cite{greywall}) is attributed to the 2D melting of a commensurate top layer crystal. It is speculated that {\em a}) such a crystal should ostensibly melt into a quasi-2D superfluid {\em b}) the intermediate ``hexatic'' phase through which melting occurs, according to the theory \cite{bers,kt}, ought to feature the superfluid properties of the fluid, giving rise to a novel ``superhexatic'' phase, possessing orientational order and capable of flowing without dissipation. \\ \indent
There are reasons to be skeptical of such an otherwise intriguing hypothesis. First, it  is highly unlikely that a single layer of $^4$He, resting on an inert solid layer, should be superfluid at a temperature as high as $T=1.4 $ K. The claims made in Ref. \cite{fukuyama} are not supported by measurements of the superfluid density; we are not aware of any monolayer $^4$He system with such a high superfluid transition temperature. Indeed, first principle computer simulations \cite {toigo} and experimental measurements \cite{vancleve}  consistently point to a superfluid transition temperature $T_c \sim$ 0.7--0.8 K for a monolayer $^4$He film on weakly attractive substrates, i.e., only slightly above that of purely 2D $^4$He \cite{worm,worm2}. It is worth noting that the failure of a {\em fluid} phase to be superfluid at a given temperature  excludes that a hypothetical {\em solid} (or orientationally ordered) phase of the same density may possess superfluid properties at that same temperature \cite{leggett}.
\\ \indent
Equally difficult to credit is the contention that a quasi-2D solid $^4$He film of density 0.075--0.080 \AA$^{-2}$ should undergo 2D melting at $T$=1.4 K, considering that 2D $^4$He at that density displays crystalline order \cite{noteb}  up to a temperature as high as 2.2 K. It is doubtful that atomic motion in the transverse direction, which acts to soften the hard core repulsion of the pair-wise helium interaction at short distances, could reduce the melting temperature by as much as $\sim 0.8$  K,  if the system is to remain essentially two-dimensional, i.e., with no significant third layer atomic promotion.  It is also worth mentioning that, while there is robust theoretical evidence that no incommensurate quasi-2D supersolid phase of $^4$He exists \cite{boninsegni11},  there is no reason to expect that a crystalline layer of $^4$He in that density range should be commensurate \cite{corboz,sm}.
\\ \indent
In order to shed light on this problem, and obtain unbiased theoretical insight into the system, we have carried out first principle QMC simulations of a thin (two layers) $^4$He film on graphite.  Our calculations are based on the standard microscopic model of $^4$He and of the graphite substrate, including accurate pair-wise interactions between $^4$He atoms, as well as between $^4$He  atoms and the graphite substrate. We present here results for the specific heat,  focusing on the coverage $\theta_A$ mentioned above, at which the specific heat ``anomaly''    is most prominent, and attempt to establish whether the standard microscopic model can account for the specific heat behavior observed experimentally, possibly supporting the contention made in Ref. \cite{fukuyama}. 
\\ \indent 
Indeed, our QMC simulations yield a very similar peak in the specific heat, but no crystalline order appears in the second layer down to the lowest temperature considered here, namely $T=0.5$ K.  Rather, the second layer is fluid-like, with no significant structural change occurring in the temperature range 0.5--1.7 K. Nor is there any evidence of a finite superfluid response at the temperature at which the peak is observed, as the layer undergoes a conventional superfluid transition at a much lower temperature, close to 0.5 K. Thus, we contend that the interpretation of the heat capacity measurements provided in Ref. \cite{fukuyama}  is unfounded, and that the data shown therein provide no support for a quasi-2D ``supersolid'' phase of $^4$He; our results point instead to atomic promotion to third layer as the most likely physical cause underlying the peak.
\\ \indent
We have also carried out simulations at a higher coverage, namely $\theta_B=0.21$ \AA$^{-2}$, at which the stable equilibrium phase of the second adlayer is an incommensurate crystal. In this case, a peak in the specific heat observed in Ref. \cite{fukuyama} at a temperature close to 1 K is claimed therein as signalling the melting of the incommensurate solid upper layer. Although we did not pursue the calculation of the specific heat for this coverage, in this case too our simulations fail to confirm the physical scenario laid out in Ref. \cite{fukuyama}, showing instead that the second layer incommensurate crystal remains stable up to a temperature of at least 2 K; moreover, melting in this system is not really ``2D'', but rather occurs through promotion of atoms to the third layer.  In summary, therefore, while the results presented here reinforce on the one hand the conclusion that no ``supersolid'' phase exists in this system, on the other they also underscore the difficulty of reliably assigning observed features in the specific heat to actual physical phenomena.
\\ \indent
The remainder of this manuscript is organized as follows: in sec. \ref{model} we describe the microscopic model adopted in this study; in sec. \ref{meth} we offer a brief description of the methodology adopted in this work; we illustrate our results in sec. \ref{res}, and outline our conclusions in sec. \ref{concl}.

\section{Model}\label{model}
 We consider an ensemble of $N$ $^4$He atoms, regarded as point-like spin-zero bosons, moving in the presence of a smooth, flat graphite substrate. 
The system is enclosed in a simulation cell shaped as a  cuboid,  with periodic boundary conditions in all directions (but the length of the cell in the $z$ direction can be considered infinite for all practical purposes). The graphite substrate occupies the $z=0$ face of the cuboid, whose area is $A$. The nominal coverage $\theta$ is given by $N/A$.
\\ \indent
The quantum-mechanical many-body Hamiltonian reads as follows:
\begin{eqnarray}\label{u}
\hat H = -\sum_{i}\lambda\nabla^2_{i}+\sum_{i}U({\bf r}_{i})+\sum_{i<j}v(r_{ij}).
\end{eqnarray}
The first and second sums run over all the $N$ $^4$He atoms,  $\lambda=6.0596$ K\AA$^{2}$, and $U$ is the potential describing the interaction of a helium atom with the graphite substrate; we use here the laterally averaged version of the well-known Carlos-Cole potential \cite{cc}. The third sum runs over all pairs of particles, $r_{ij}\equiv |{\bf r}_i-{\bf r}_j|$ and $v(r)$ is the accepted Aziz pair potential \cite{aziz}, which describes the interaction between two helium atoms. 
\\ \indent
As mentioned above, this is the standard microscopic model used to describe a thin helium film on a substrate. The substrate itself is considered smooth and flat, i.e., its  corrugation is neglected (as well as, obviously, zero-point motion of the carbon atoms in the substrate). This has been quantitatively shown to be a valid approximation; for example, the effect of explicitly including the corrugation on the computed energy difference between solid and liquid phases is of the order of 0.01 K \cite{sm}. One can understand why that is the case, by considering that the lower $^4$He solid layer is incommensurate with the graphite substrate, while the upper layer mostly experiences the corrugation of the lower layer.
\\ \indent
At the $^4$He coverages and in the temperature range considered here, two atomic layers form.  
In principle, of course, $^4$He atoms are identical, and therefore no conceptual distinction can be drawn between atoms in the ``top'' and ``bottom'' layer. However, we have carried out a number of simulations of the system with all atoms regarded as indistinguishable, consistently observing  that both inter-layer hopping of atoms, as well as quantum-mechanical exchanges among atoms in the first adsorbed  layer (which orders as a triangular crystal) and/or in different layers, are {\em exceedingly infrequent}. It is therefore an  excellent approximation to regard atoms in the bottom layer as {\em distinguishable} quantum particles (i.e., ``Boltzmannons''); on the other hand, atoms in the top layer are considered as indistinguishable, and can therefore undergo quantum exchanges. 
\\ \indent
As a consequence of the above approximation (one that is routinely made in computer simulations of this system), which underlies all of the results that we present here, the numbers of atoms in the two layers are constant, i.e., $N_1$ atoms constitute the first layer, $N_2=N-N_1$ the second (and $N_1/A$, $N_2/A$ are the 2D densities in the two layers).  This {\em de facto} amounts to regarding the two layers as two distinct ``species'', which  allows us to compute separately their energetic contributions. In turn, this makes it possible for us to test quantitatively the contention made in Ref. \cite{fukuyama}, namely that the bulk of the contribution to the specific heat of the film comes from the top layer.
\section{Methodology}
\label{meth}
We carried out QMC simulations of the system described in section \ref{model} using the worm algorithm in the continuous space path integral representation \cite{worm,worm2}. We shall not review the details of this method, referring instead the reader to the original references. We utilized a canonical variant of the algorithm in which the total number of particles $N$ is held constant, in order to simulate the system at fixed coverage \cite{mezz1,mezz2}.
We obtained results in a range of temperature going from 0.5 K to a high temperature of 1.7 K (3 K) for coverage $\theta_A \ (\theta_B)$. The number $N_1$ of $^4$He atoms in the lower level is either 64 or 144, whereas that ($N_2$) of atoms in the upper layer is determined by the total $^4$He coverage, as well as the 2D density of the lower level (we take that from experiment, whenever available). For a discussion of the dependence of the results on the size of the simulated system, we refer the reader to the Appendix.
\\ \indent
Details of the simulation are standard; we made use of the fourth-order approximation for the high-temperature density matrix (see, for instance, Ref. \cite{boninsegni05}), and all of the results quoted here are extrapolated to the limit of time step $\tau\to 0$. In general, we found that a value of the time step equal to $1.6\times 10^{-3}$ K$^{-1}$ yields estimates indistinguishable from the extrapolated ones.
\\ \indent
The main physical quantity of interest is the specific heat $C(T)$, which we aim at comparing with that measured experimentally. It is well known that a direct calculation of the specific heat in QMC simulations is complicated by the fact that estimators are statistically ``noisy''. It is therefore easier to obtain $C(T)$ by computing the energy per particle $e(T)$ and obtaining $C(T)$ as $de(T)/dT$, either through numerical differentiation or by  fitting the computed $e(T)$ curve.\\
The occurrence of crystalline order in the system is detected through i) visual inspection of the imaginary-time paths and ii) the calculation of the pair-correlation function $g(r)$, integrated along the direction ($z$) perpendicular to the substrate. Superfluid order is detected through the direct calculation of the superfluid fraction, using the well-established winding number estimator \cite{pollock}. Qualitative insight on the propensity of the system to flow without dissipation can be  obtained from the computed statistics of exchange cycles.

\section{Results}\label{res}
As mentioned in the Introduction, we have computed the specific heat as a function of temperature for the single coverage $\theta_A=0.1973$ \AA$^{-2}$ for which the peak  observed in Ref. \cite{fukuyama} is strongest and occurs at the highest temperature (close to 1.4 K). We set the density of the lower layer to 0.1205 \AA$^{-2}$, as specified in Ref. \cite{fukuyama}.
\begin{figure}[h]
\centering
\includegraphics[width=0.47\linewidth]{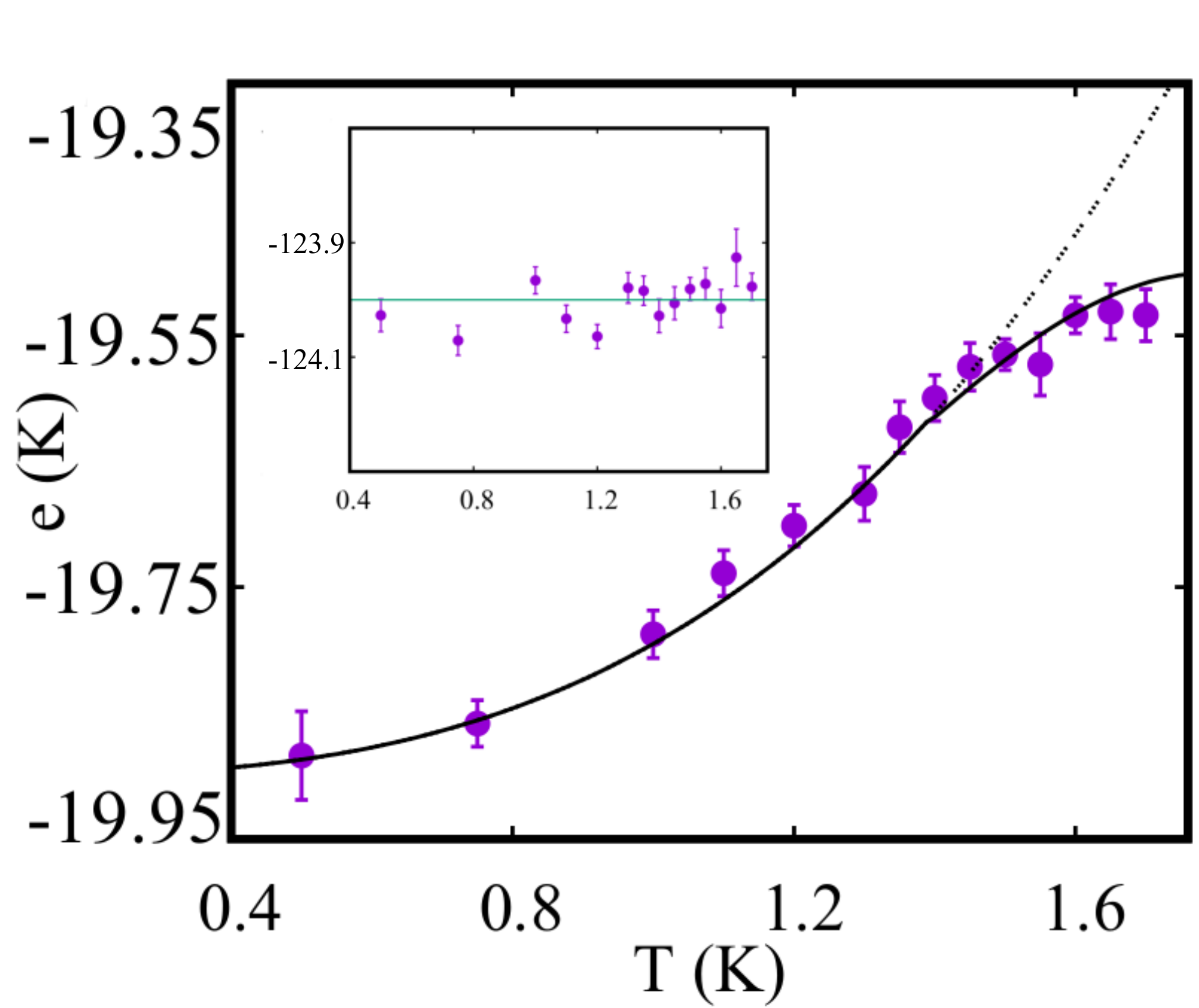}
\includegraphics[width=0.48\linewidth]{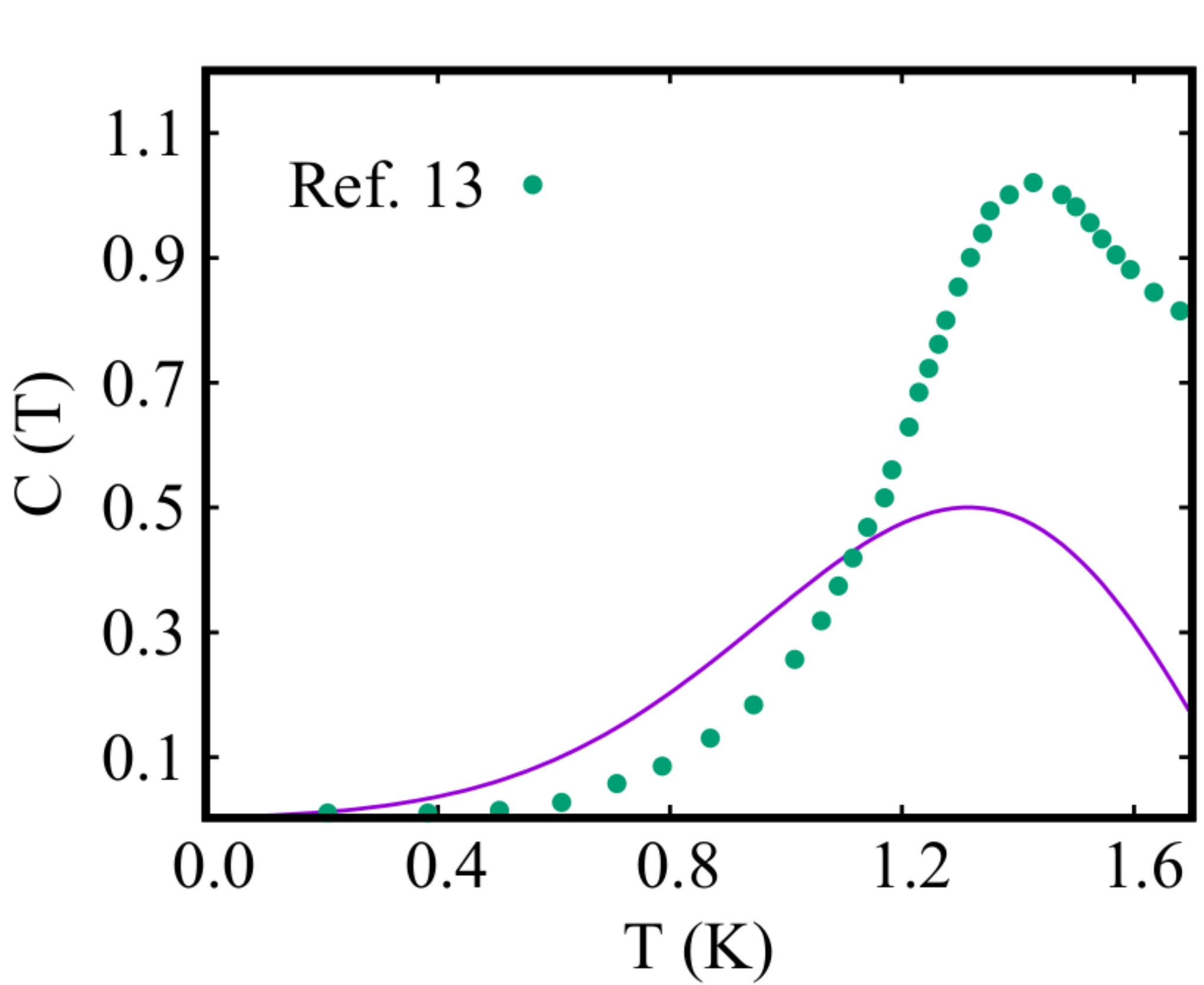}
\caption{{\em Left}: Energy per $^4$He atom (in K) in the top layer, as a function of temperature.
Solid line is a fit to the data. Dashed line is a cubic fit to the data for $T < 1.4$ K. Inset shows the energy per particle for the bottom layer. These results are obtained by simulating a system comprising $N_1=64$ ($N_2=41$) particles in the bottom (top) layer. {\em Right}: Specific heat of the film, estimated by differentiating the function used to fit the data for $e(T)$ for the top layer. 
%MB 1125
%The fitting curve has a statistical uncertainty worth approximately 5\% of the values plotted, arising from the uncertainty in %the determination of the fitting parameters. 
%MB end 
Also shown 
%MB 1125 for comparison 
are experimental measurements from Ref. \cite{fukuyama}.}
\label{f1}
\end{figure}

Fig. \ref{f1} (left) shows the computed energy per $^4$He atom $e(T)$ as a function of temperature, for the top layer. The inset shows the corresponding quantity for the bottom layer. The first observation is that there is no noticeable dependence on the temperature of the estimates for the bottom layer, showing that the contribution to the specific heat of the film comes almost exclusively from the top layer, as opined in Ref. \cite{fukuyama}. The energy values for the top layer follow the expected (phonon) $\sim T^3$ behavior at low $T$ (dashed line in Fig. \ref{f1}), but we cannot fit all of our data with a single power-law expression, as there is a clear inflexion, resulting in a broad peak of the specific heat $C(T)\equiv de(T)/dT$ at $T\sim 1.3$ K. This is shown in the right panel of Fig. \ref{f1}, displaying the derivative with respect to the temperature of the fitting curve in the left panel (solid line). Also shown 
%MB 1125
%for comparison 
%MB 1125
are the experimental data for the specific heat of Ref. \cite{fukuyama}, read off Fig. 1(c) therein.
\\ \indent
The comparison of our results with experiment appears altogether satisfactory, in that the most important features of the experimentally measured specific heat, i.e., the presence of a peak, its overall shape and the temperature at which it occurs, are all fairly well reproduced. 
Achieving a quantitative reproduction of the experimental measurements is beyond the scope of this study, as the microscopic model utilized here, arguably one of the most reliable presently available, is nonetheless still relatively simplified (e.g., the graphite substrate is regarded as flat), and based on semi-empirical potentials.
%MB 1125
It should also be mentioned that carrying out a detailed comparison of  the specific heat  experimentally measured in Ref. \cite{fukuyama}, and that yielded by our energy estimates is no trivial task, as {\em a}) a straightforward numerical differentiation of the set of computed energy values shown in Fig. \ref{f1} is affected by a relatively large uncertainty, and {\em b}) there is ambiguity associated with the fitting procedure, as different functional forms can fit the data while yielding significantly different $C(T)$ curves. Nevertheless, the most important physical result, namely the presence of a peak at experimentally relevant temperatures, is robust, and does not require that the underlying model be microscopically accurate. This gives us confidence in our ability to assess the plausibility of the physical scenario proposed in Ref. \cite{fukuyama} to account for such a peak.
\begin{figure}[t]
\centering
\includegraphics[width=1.0\linewidth]{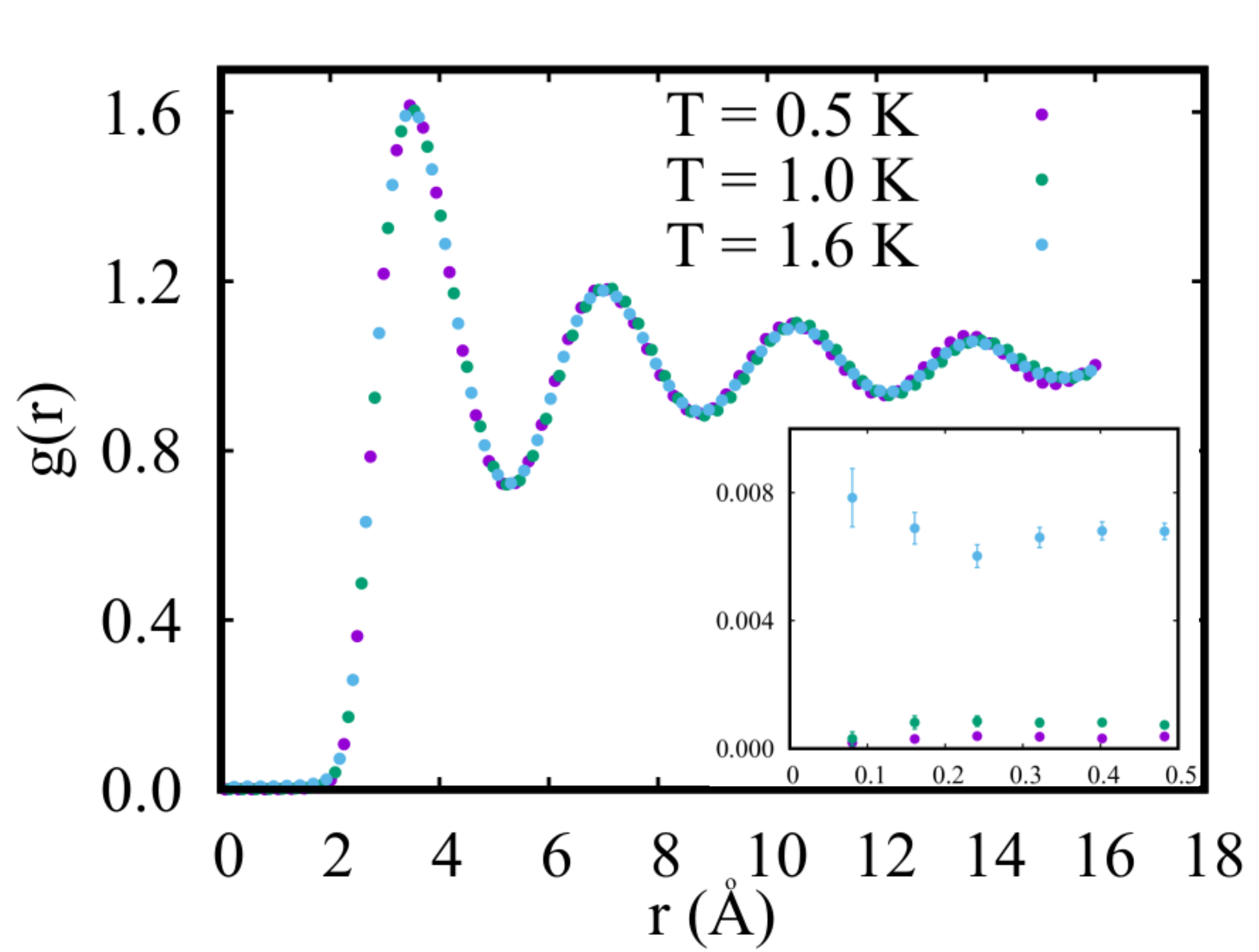}
\caption{Pair correlation function $g(r)$ for the upper layer of a $^4$He film of coverage $\theta_A=0.1973$ \AA$^{-2}$, integrated over the axis perpendicular to the substrate. These results shown here for three different temperatures ($T$ =0.5, 1.0 and 1.6 K) are obtained by simulating a system comprising $N_1= 144\ (N_2= 92)$ particles in the bottom (top) layer.  Inset shows a blow-up of the region near $r=0$. When not shown, statistical uncertainties are smaller than the symbol size.}
\label{f3}
\end{figure}
\\ \indent We begin by examining possible structural changes occurring in the film as the temperature is lowered from $T=1.6$ K through $T=0.5$ K, i.e., across the specific heat anomaly, occurring in our case at $T\sim$ 1.3 K, i.e., a slightly lower temperature than in the experiment. Fig. \ref{f3} shows the reduced pair correlation function $g(r)$ for the top layer (of 2D density 0.0768 \AA$^{-2}$), integrated over the direction perpendicular to the substrate. The results clearly show little or no difference among the $g(r)$ computed at these three significantly different temperature, pointing to the absence of structural change in the film. The rapidly decaying oscillations indicate that the system is in the liquid phase, as confirmed by visual inspection of many-particle configurations generated by the random walk. 
The only, rather subtle change that takes place as the temperature is raised, is illustrated in the inset of Fig. \ref{f3}, in which the region near $r=0$ is shown magnified. The fact that $g(r)$ remains finite in the $r\to 0$ limit, at the highest temperature, is evidence of promotion of atoms to the third layer, which, in our submission, constitutes the most plausible physical explanation for the peak in the specific heat observed both in our study and in the experiment.
Specifically, we attribute the change of slope of the $e(T)$ curve (Fig. \ref{f1}), giving rise to the peak in $C(T)$, to the reduced second layer density arising from atomic promotion to the third layer. The results obtained in this study are entirely consistent with the physical picture offered in, e.g., Ref. \cite{corboz}, i.e., they do {\em not} support the contention that such an anomaly should arise from the melting of a 2D crystal, as contended in Ref. \cite{fukuyama}.
\\ \indent
Nor is there any evidence of a finite superfluid response of the second layer (as mentioned above, the bottom layer is crystalline and inert) for $T > 1$ K. Although permutations of indistinguishable particles do occur, even at temperatures as high as 1.7 K, the expected superfluid transition of the fluid layer takes place below 1 K. While we did not pursue the precise determination of $T_c$ in this work, the values of the superfluid fraction obtained on a system of 92 particles in the top layer suggest that $T_c\sim 0.5$ K (the superfluid fraction approaches 100\% at $T=0.25$ K). In other words, no superfluid phase of the system exists at temperatures as high as those at which the specific heat anomaly reported in Ref. \cite{fukuyama} occurs. In light of this, the interpretation offered in Ref. \cite{fukuyama} of the specific heat measurements seems unviable.
\\ \indent 
In order to illustrate how problematic it is to attribute specific physical meanings to distinct features of the measured specific heat, we turn now to a higher coverage, namely $\theta_B=0.21$ \AA$^{-2}$, for which the specific heat displays a sharp peak at a temperature $T$ between 1.1 and 1.2 K. This peak is interpreted in Ref. \cite{fukuyama} as indicative of the 2D melting of an {\em incommensurate} crystal. There is no controversy here, as to whether at that coverage the second layer should form an incommensurate solid; but, as we show below, neither the contention that such a layer should undergo melting at a temperature so low, nor  that the melting transition should be 2D in nature, are  supported by our first principle simulations.
\begin{figure}[t]
\centering
\includegraphics[width=1.0\linewidth]{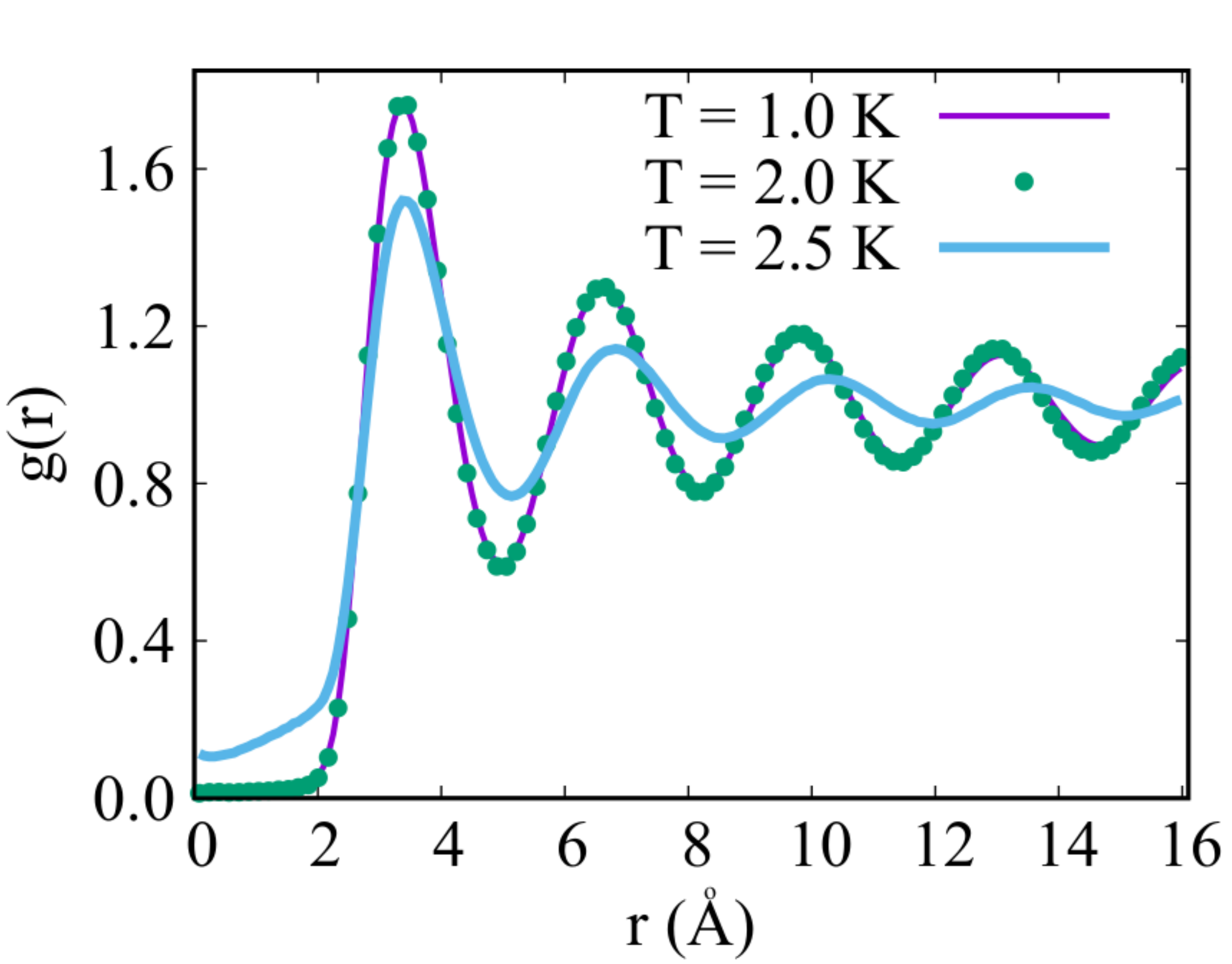}
\caption{Pair correlation function $g(r)$ for the upper layer of a $^4$He film of coverage $\theta_B=0.21$ \AA$^{-2}$, integrated over the axis perpendicular to the substrate. These results shown here for three different temperatures ($T=$ 1.0, 2.0 and 2.5 K) are obtained by simulating a system comprising $N_1= 144\ (N_2= 106)$ particles in the bottom (top) layer. The density of the first layer is fixed at 0.1209 \AA$^{-2}$. }
\label{f4}
\end{figure}
\\ \indent
Fig. \ref {f4} shows the same quantity as in Fig. \ref{f3}, namely the reduced pair correlation function for the top layer, for three different temperatures, namely $T$ = 1.0, 2.0 and 2.5 K. Here, we  set the value of the density for the bottom layer (not given in Ref. \cite{fukuyama}) to be 0.1209 \AA$^{-2}$. It is manifest from the data shown in the figure that virtually {\em nothing} happens to the film, structurally, between 1 and 2 K, and in particular there is no discernible attenuation of the oscillations that mark the presence of crystalline long-range order. On the other hand, at $T=2.5$ K such oscillations decay rapidly, the main peak has lost $\sim$ 20\% of the strength and, most significantly, $g(0)$ is finite and relatively large, signalling that the disappearance of order (i.e., melting) is connected to the promotion of atoms to the third layer, i.e., it cannot be regarded as ``two-dimensional". We have not pursued the fairly lengthy calculation of the specific heat for this coverage, but it is clear that, in this case too, the interpretation of the measurements proposed in Ref. \cite{fukuyama} clashes with the results of our computer simulations.

\section{Conclusion}\label{concl}

We have carried out extensive, first principle numerical study of structure and energetics of a two-layer $^4$He film adsorbed on graphite, with the aim of gaining insight into the physics of the system, in light of recent experimental measurements for which a potentially exciting interpretation was put forth. We have made use of state-of-the-art numerical techniques (QMC); the quantitative predictive power of this methodology, for interacting Bose systems, is by now fairly well-established. As well, the microscopic model utilized here, based on accepted potentials, has been adopted in essentially {\em all} previous numerical studies \cite{pipo}, including those based on QMC simulations, which have yielded predictions generally in agreement with experiment.
\\ \indent
It seems fair to state that even in this case there is substantial agreement between theoretical results and experimental data; quantitative differences can be attributed to the inevitable limitations of a microscopic model which, while capturing the bulk of the physical effects, is nonetheless still highly simplified. 
However, our results fail to provide any kind of support for the interpretation of experimental data proposed in Ref. \cite{fukuyama},  even though they  confirm some of the working assumptions made therein, e.g., that the contribution to the specific heat comes almost entirely from the second layer. In particular, we see no evidence of melting of a crystal (either commensurate or incommensurate) at the temperatures at which anomalies in the specific heat are observed experimentally, nor is there any evidence of superfluid behaviour where, according to the authors of Ref. \cite{fukuyama}, a ``superfluid hexatic'' phase may occur. More generally, the data presented in Ref. \cite{fukuyama} offer nothing to the effect that the theoretical phase diagram of this system, as proposed for example in Ref. \cite{corboz}, should be substantially revised.
\\ \indent
It is important to restate at this point that no direct experimental evidence has been produced so far of a commensurate solid phase in the second layer of $^4$He on graphite. Its existence has so far been only posited, as a plausible way to account for specific features in the experimentally observed specific heat. But, as also shown in this work, the interpretation of those features is often not univocal; for example, there are valid reasons to attribute the broad peak in the specific heat at the lower coverage investigated here, to promotion of atoms to the third layer; while the physical nature of the peak in the specific heat at the higher coverage was not investigated here, there is no reason to exclude such an explanation in that case as well. 
\\ \indent
In summary, it seems as if progress toward the resolution of the controversy existing at the moment, over the presence of a commensurate phase in this system, is not likely to be achieved through measurements of the specific heat, whose interpretation is ambiguous. It seems as if, at this point, alternate sources of experimental information are needed, ideally capable of directly imaging the second layer and  providing robust evidence for the existence of the elusive commensurate phase.
\appendix*
\section{\label{sec:Aone} Dependence of results of system size}
\begin{figure}[t]
\centering
\includegraphics[width=1.0\linewidth]{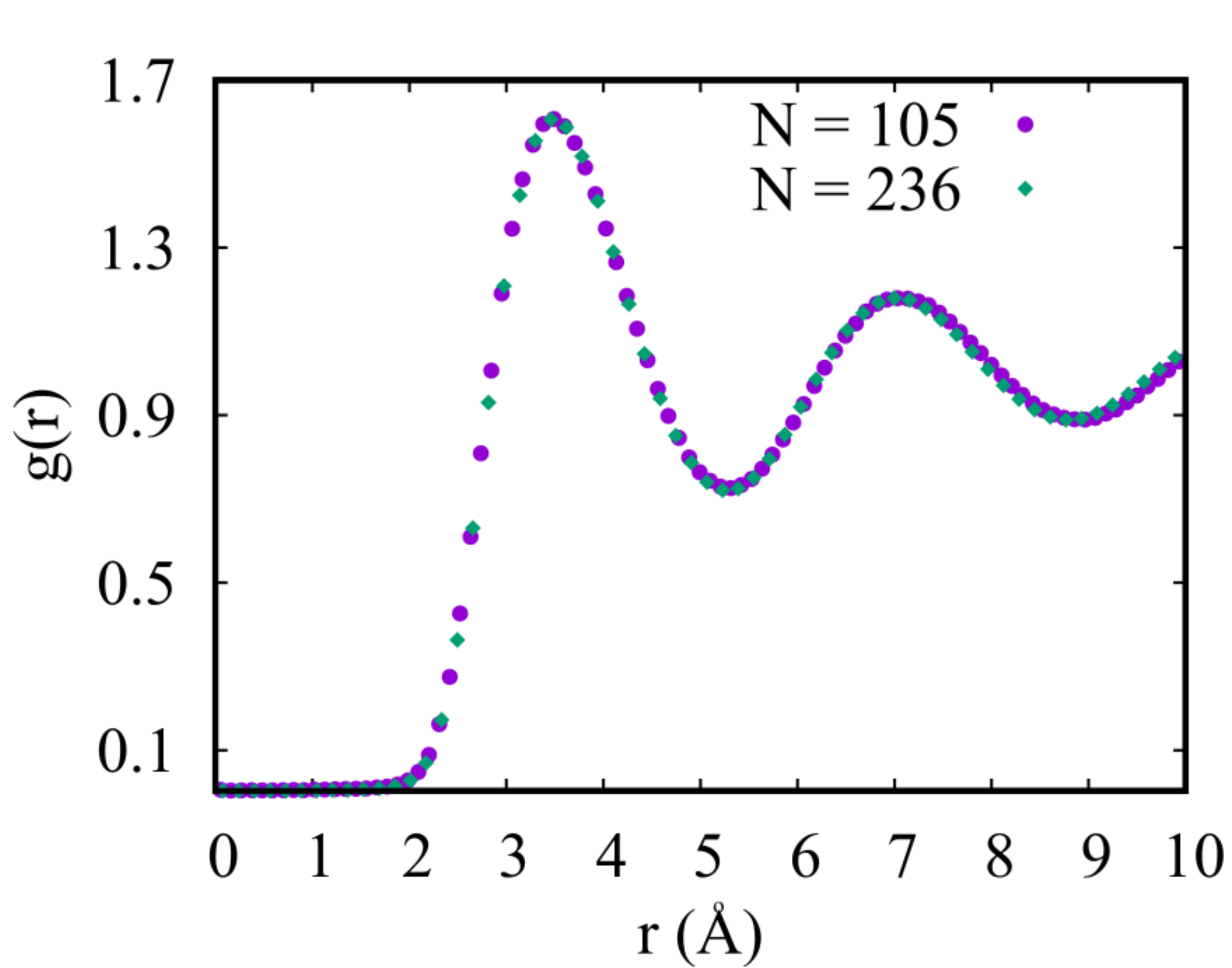}
\caption{Pair correlation function $g(r)$ for the upper layer of a $^4$He film of coverage $\theta_A=0.1973$ \AA$^{-2}$, integrated over the axis perpendicular to the substrate. These results shown here are for temperature $T=1.4$ K, and are obtained by simulating a system comprising $N_1=64\ (N_2=41)$ $^4$He atoms in the bottom (top) layer (circles), as well as $N_1= 144\ (N_2= 92)$ $^4$He atoms in the bottom (top) layer (diamonds). The density of the first layer is fixed at 0.1205 \AA$^{-2}$. Statistical errors are smaller than symbol sizes.}
\label{ffs}
\end{figure}

We offer here some details on the dependence on system size of the physical estimates obtained in our Monte Carlo simulation, by providing a few explicit, representative results for the two system sizes considered here. 
\begin{table}[h]\label{t1}
\begin{ruledtabular}
\begin{tabular}{cccc}
$T\ (K)$ & $N$ & $N_2$ & $e\ (K)$ \\ \hline
0.5 & 105 & 41 & $-19.877\ (35)$ \\
0.5 & 236 & 92 &$-19.847 \ (34)$\\
1.4 & 105 & 41 & $-19.593\ (18)$ \\
1.4 &236 & 92 & $-19.565\ (29)$\\
\end{tabular}
\caption{Energy per $^4$He atom in the upper layer of a film of coverage $\theta_A =0.1973$ \AA$^{-2}$, computed at two different temperatures for the two system sizes considered in this work. Statistical errors (in parentheses) are on the last two digits.}
\end{ruledtabular}
\end{table}

Table I reports computed energy per $^4$He atom in the top layer of a film of coverage $\theta_A=0.1973$  \AA$^{-2}$. The estimates are  at two different temperatures, and show consistency of the results, within the statistical errors of the calculation.  This is not a surprise, as it is generally accepted that the energy is not particularly sensitive to the size of the simulated system, as it is mostly affected by the environment experienced by atoms in their immediate vicinity. In all cases, the contribution to the potential energy arising from interatomic distances greater than $r_c=10.7$ \AA\ is evaluated by assuming atomically thin, flat layers, and setting $g(r)=1$ beyond $r_c$.
Fig. \ref{ffs} shows the integrated pair correlation function $g(r)$ defined above, for a film of coverage $\theta_A$, at a temperature $T=1.4$ K.  Consistency of the results yielded by the two different sizes is clear.

\section*{Acknowledgements}
One of us (MB) acknowledges the hospitality of the International Centre for  Theoretical Physics, Trieste (Italy), where part of this research work was carried out.
This work was supported in part by the Natural Sciences and Engineering Research Council of Canada (NSERC). Computing support of ComputeCanada is gratefully acknowledged.

% SECOND OPTION:
% Use your bibtex library
% \bibliographystyle{SciPost_bibstyle} % Include this style file here only if you are not using our template
%\bibliography{SciPost_Example_BiBTeX_File.bib}

\begin{thebibliography}{99}
\bibitem{noteb}
 This conclusion was established in this work by means of targeted Quantum Monte Carlo simulations of two-dimensional $^4$He.
\end{thebibliography}

\end{document}